\documentclass[final,1p,times]{elsarticle}
\newcounter{bla}

\journal{Computer Physics Communications}
\usepackage[utf8]{inputenc}
\usepackage[T1]{fontenc}
\usepackage{fixltx2e}
\usepackage{graphicx}
\usepackage{longtable}
\usepackage{float}
\usepackage{wrapfig}
\usepackage{rotating}
\usepackage[normalem]{ulem}
\usepackage{amsmath}
\usepackage{textcomp}
\usepackage{marvosym}
\usepackage{wasysym}
\usepackage{amssymb}
\usepackage{hyperref}
\usepackage{amsmath}
\usepackage{libertine}
\newcommand{\ZZ}{\ensuremath{\mathbb{Z}}}
\newcommand{\Fp}{\ensuremath{\mathbb{F}_p}}
\usepackage{algorithm}
\usepackage[noend]{algpseudocode}
\algrenewcommand\algorithmicrequire{\textbf{Input:}}
\algrenewcommand\algorithmicensure{\textbf{Output:}}
\DeclareMathOperator{\rank}{rank}
\date{\today}
\title{Finding Linear Dependencies in Integration-By-Parts Equations: A Monte Carlo Approach}
\hypersetup{
  pdfkeywords={},
  pdfsubject={},
  pdfcreator={Emacs 24.3.1 (Org mode 8.2.1)}}
\begin{document}

\begin{frontmatter}
\author[a]{Philipp Kant}
\address[a]{Humboldt-Universität zu Berlin, Institut für Physik, Newtonstraße 15, 12489 Berlin}
\ead{philipp.kant@physik.hu-berlin.de}
\begin{abstract}
The reduction of a large number of scalar integrals to a small set of
master integrals via Laporta's algorithm is common practice in
multi-loop calculations.  It is also a major bottleneck in terms of
running time and memory consumption.  It involves solving a large set
of linear equations where many of the equations are linearly
dependent.  We propose a simple algorithm that eliminates all linearly
dependent equations from a given system, reducing the time and space
requirements of a subsequent run of Laporta's algorithm.
\end{abstract}
\end{frontmatter}
{\bf PROGRAM SUMMARY}

\begin{small}
\noindent
{\em Program Title\/:} ICE---the IBP Chooser of Equations\\
{\em Available From\/:} \href{http://www.physik.hu-berlin.de/pep/tools}{http://www.physik.hu-berlin.de/pep/tools}\\
{\em Programming language\/:} Haskell\\
{\em Computer\/:} any system that hosts the Haskell Platform                                              \\
{\em Operating system\/:} GNU/Linux, Windows, OS/X                
{\em Classification\/:} 4.4, 4.8, 5, 11.1                            \\
{\em Nature of problem\/:}
  find linear dependencies in a system of linear equations with multivariate polynomial coefficients.  To be used on Integration-By-Parts identities before running Laporta's Algorithm.
   \\
{\em Solution method\/:}
  map the system to a finite field and solve there, keeping track of the required equations.
   \\
{\em Restrictions\/:}
  typically less than the restrictions imposed by the requirement of being able to process the output with Laporta's Algorithm.
   \\
{\em Unusual features\/:}
  complexity increases only very mildly with the number of kinematic invariants.
   \\
{\em Running time\/:}
  depends on the individual problem.  Fractions of a second to a few minutes have been observed in tests.
   \\
\end{small}

\section{Introduction}
\label{sec-1}
In multi-loop calculations, one often finds that the expression for a
given Feynman diagram, after tensor decomposition, is given in terms
of a very large number of integrals of the form
\begin{equation}
I(\nu_1,\dots,\nu_n) =
\int d^dk_1 \cdots d^dk_l
\frac{1}{D_1^{\nu_1} \cdots D_n^{\nu_n}}
\,.
\label{eq:1}
\end{equation}
Here, $\nu_i \in \ZZ$ are called the \emph{indices} of a given integral.
The $D_i$ are polynomials of total degree 2 in the loop momenta $k_i$
and any external momenta and masses.  Integrals with different indices
satisfy a set of linear relations, and it is desirable to express a
diagram using a minimal set of linearly independent integrals, the
so-called \emph{master integrals}.

One source of linear equations relating different integrals are the
\emph{Integration-By-Parts} (IBP) identities
of~\cite{Tkachov:1981wb,Chetyrkin:1981qh}.  They are a consequence of
translational invariance of the integral.  Additional relations are
obtained from Lorentz invariance (LI)~\cite{Gehrmann:1999as}.
Both IBP and LI equations relate an integral with indices $\{\nu_i\}$
to integrals where some of the $\{\nu_i\}$ are shifted.  The
coefficients are multivariate polynomials of total degree at most 1 in
scalar products of external momenta, squared masses, and the
space-time dimension $d$.

Laporta~\cite{Laporta:2001dd} has given an algorithm that
systematically solves IBP and LI identities to reduce a given set of
integrals to a linearly independent set.  Underlying the algorithm is
the observation that allowing larger indices, the number of integrals
grows slower than the number of IBP and LI identities relating these
integrals with each other.  At some point, the rank $r$ of the system
is sufficiently large that all integrals within a certain range of
indices can be reduced to a small number of master integrals.

Laporta's algorithm proceeds by defining an order on the set of
integrals that corresponds roughly to the difficulty of calculating
them.  In each step of the algorithm, one equation is solved for the
most "difficult" integral, and the equations solved in earlier steps
are inserted.  Finally, all integrals are expressed through a set of
"simple" master integrals.  The algorithm has become a standard
procedure in higher order calculations, and several public
implementations~\cite{Anastasiou:2004vj,Smirnov:2008iw,Studerus:2009ye,vonManteuffel:2012np,Smirnov:2013dia}
are available.

There are two inconveniences that cause Laporta's algorithm to be
resource hungry.  One is intermediate expression swell: starting with
polynomial coefficients of low degree, the process of solving and
substituting leads to equations over rational functions of high
degree, and with large coefficients.  Intermediate expressions are
usually much larger than the final answer and can challenge the
available memory and disk space.  In order to mitigate the growth of
coefficients and minimise memory usage, the intermediate expressions
are regularly simplified, so that the overall running time of the
algorithm is dominated by multivariate $\gcd$ calculations and rational
function simplification.

This build-up of large intermediate expressions is amplified by the
second problem: the number of IBP and LI equations relating a given
set of integrals is much larger than their rank, the number of
equations that are linearly independent.  Consequently, much time is
spent processing redundant information, effectively calculating a lot
of zeros.  Eliminating the redundancy in the linear system has the
potential to reduce the demands on CPU time and memory.

The problem of identifying linearly dependent equations beforehand has
seen some investigation.  For instance, Lee~\cite{Lee:2008tj}
gives selection criteria based on the group structure of the IBP and
LI identities.
We follow a different approach and propose an algorithm that detects
linear dependencies in a given set of IBP and LI equations, thus
reducing the time and space requirements of a subsequent run of
Laporta's algorithm.  Our algorithm is randomised in the Monte Carlo
sense, i.e., it has deterministic running time and gives the correct
answer with high probability.
\section{The Algorithm}
\label{sec-2}
We now present an algorithm that removes any redundant equations from
a system of linear equations with multivariate polynomial
coefficients.  In the case of Laporta's algorithm, this can
drastically reduce the size of the system, and thus the required CPU
time and memory.

The basic idea is this: writing the system in matrix form, where each
column corresponds to one integral and each row to a linear
relationship between integrals, and solving by Gaussian elimination
would reduce linearly dependent rows to zero during the forward
elimination, allowing the identification and removal of redundant
equations.  But there would be no gain: determining the minimal set of
equations would require the solution of the whole system in the first
place.

However, the cost of Gaussian elimination can be reduced by mapping
the coefficients homomorphically to a simpler domain.  As long as the
homomorphism does not reduce the rank of the system, one can still
read off which equations are redundant.  We follow the canonical
choice of using $\Fp$, the field of integer numbers modulo a prime
$p$.  In this way, the Gauss algorithm does not suffer from
intermediate expression swell, and no $\gcd$ calculations are
necessary\footnote{A pedagogical introduction to the technique of homomorphic
images can be found, for example,
in~\cite{zippel1993effective,Gathen:2003:MCA:945759}.}.

\begin{algorithm}
\caption{Get a maximal linearly independent subset of a given system of linear equations over $\ZZ[x_1,\dots,x_s]$.
\label{alg:findlindep}}
\begin{algorithmic}[1]
\Require $A$, an $n\times m$ matrix over $\ZZ[x_1,\dots,x_s]$.
\Ensure $B$, an $r\times m$ submatrix of $A$ with linearly independent rows, where $r \leq \rank A$.  With high probability, $r = \rank A$.
\Statex
\State $p \gets$ a large prime number
\State\label{step:mod} $A' \gets A \mod p \in \left(\Fp[x_1,\dots,x_s]\right)^{n\times m}$
\Statex\Comment{Take the residue $\mod p$ of every coefficient of every polynomial.}
\State $a_1,\dots,a_s \gets$ random points from \Fp
\State\label{step:eval} $A'' \gets A'(a_1,\dots,a_s) \in \Fp^{n\times m}$
\Statex\Comment{Evaluate every entry of $A'$ at the point $(x_1=a_1,\dots,x_s=a_s) \mod p$.}
\State\label{step:gauss} Perform forward Gauss elimination on $A''$.
  Before each step, perform a row permutation to get a non-zero pivot element.  
  Let $I=\{i_1,\dots,i_n\}$ be the resulting permutation of rows, and $r$ the number of non-zero rows after Gaussian elimination (i.e., the rank of $A''$).
\State $B \gets$ the matrix consisting of rows $i_1,\dots,i_r$ of $A$
\end{algorithmic}
\end{algorithm}

The resulting algorithm is depicted in
Algorithm~\ref{alg:findlindep}.  The operation of taking the
modulus of $A$ in step~\ref{step:mod} is meant to be element-wise:
we take the modulus of each coefficient of each polynomial in the
matrix.  Likewise, the evaluation of the matrix in
step~\ref{step:eval} is meant as an evaluation (within \Fp) of
every polynomial.

It should be noted that in addition to identifying a maximal linearly
independent set of equations, the algorithm also
identifies the master integrals: any column that does not contain a
pivot element corresponds to an integral that cannot be reduced with
the given set of equations.  Of these, some will be integrals with
large indices that could be solved with additional equations, and the
others will be the master integrals.

\subsection{Simple Example}
\label{sec-2-1}
In order to illustrate the algorithm, we give a simple example.
Consider
\begin{equation*}
A=\left(\begin{matrix}
x & x+y & 1 & 0\\
5x & 3y & 0 & x\\
-4x & x-2y & 1 & -x\\
0 & x & y & 3x\\
x & 2x + y & y+1 & 3x
\end{matrix}\right)
\in {\ZZ[x,y]}^{5\times 4}
\,.
\end{equation*}
Choosing \(p=29\) and evaluating at \(x=6, y=26\) yields
\begin{equation*}
A''=\left(
\begin{matrix}
6 & 3 & 1 & 0 \\
1 & 20 & 0 & 6 \\
5 & 12 & 1 & 23 \\
0 & 6 & 26 & 18 \\
6 & 9 & 27 & 18 \\
\end{matrix}\right)
\in \mathbb{F}_{29}^{5\times 4}
\,.
\end{equation*}
Performing Gaussian forward elimination, we get
\begin{align*}
&\left(\begin{matrix}
6 & 3 & 1 & 0 \\
0 & 5 & 24 & 6 \\
0 & 24 & 5 & 23 \\
0 & 6 & 26 & 18 \\
0 & 6 & 26 & 18 \\
\end{matrix}\right)
&&\to&&
\left(\begin{matrix}
6 & 3 & 1 & 0 \\
0 & 5 & 24 & 6 \\
0 & 0 & 0 & 0 \\
0 & 0 & 3 & 5 \\
0 & 0 & 3 & 5 \\
\end{matrix}\right)
&&\to&&
\left(\begin{matrix}
6 & 3 & 1 & 0 \\
0 & 5 & 24 & 6 \\
0 & 0 & 3 & 5 \\
0 & 0 & 0 & 0 \\
0 & 0 & 0 & 0 \\
\end{matrix}\right)\,
\end{align*}
Before the last step, we had to exchange the third and fourth row.
This tells us that the first, second and fourth rows of \(A\) are
linearly independent, and we return
\begin{equation*}
B=\left(
\begin{matrix}
x & x+y & 1 & 0\\
5x & 3y & 0 & x\\
0 & x & y & 3x\\
\end{matrix}\right)
\in {\ZZ[x,y]}^{3\times 4}
\,.
\end{equation*}
\subsection{Probability of Failure}
\label{sec-2-2}
The algorithm will always return $r$ linearly
independent rows of $A$.  This follows because the rows
$i_1,\dots,i_r$ of $A''$ are linearly independent by construction,
and $A''$ is obtained from $A$ by evaluation and taking the residue
modulo $p$, which cannot remove a linear dependency.
In unlucky cases, however, the rank can be decreased while going from
$A$ to $A''$.  In this case, we erroneously discard too many equations
and wind up with more master integrals than necessary.  
It is rather unlikely that this actually happens, and we can give an
upper bound on the probability of running into such an unlucky case.

In order to derive this bound, let us consider a modification of the
algorithm: instead of performing Gaussian elimination on $A''$, we
perform fraction-free Gaussian elimination on the original matrix $A$,
and map the result to $\Fp^{n\times m}$.
The rank of the resulting matrix can only be reduced if one of the
pivot elements is mapped to zero by evaluation at $a_1,\dots,a_n$ or
by taking the residue $\mod p$.  A bound on the probability for this
can be derived using the Schwartz-Zippel
lemma~\cite{Demillo1978193,zippel1979,schwartz1980fast}: the value
of a non-zero polynomial of total degree $d$ at a point taken randomly
from a set of cardinality $p$ is zero with probability at most
$\tfrac{d}{p}$.

The degree of the \(i\)th pivot element in a fraction-free Gaussian
elimination is bounded by \(i\delta\), where $\delta$ is the maximal
degree of entries in the initial matrix (see, for
example~\cite{GVK110761715}).  
In the case of IBP equations, the entries of the matrix \(A\) are of
degree at most one, so \(\delta=1\), and the probability that none of
the pivot elements are accidentally evaluated at a zero is at least
\begin{equation}
\label{eq:2}
P(\text{success}) \geq
\prod_{i=1}^r\left(
1 - \frac{i}{p}\right)\,.
\end{equation}

In order to transfer this bound to our algorithm,
where we take the residue and evaluate before the elimination step, we
have to make sure that these homomorphisms commute with the operation
of performing Gaussian elimination.  Generally, this is not the case.
However, if a homomorphism does not map any of the pivot elements to
zero --- which is exactly the case we considered --- that homomorphism
does commute with the elimination
step~\cite{McClellan:1973:ESS:321784.321787}.

By choosing a large prime $p$ in~(\ref{eq:2}), we can get a high
probability of finding the maximal set of linearly independent
equations.  Of course, machine restrictions may prevent us from
choosing arbitrarily large primes, so for very large systems we might
not be able to get~(\ref{eq:2}) as large as we would want.  In
such a case, it is always possible to run the algorithm multiple
times, keeping the result with the maximal rank, decreasing the
probability of failure exponentially.

The estimate (\ref{eq:2}) seems to be a rather conservative bound.  In
testing the algorithm, we have deliberately used rather small primes
in order to increase the calculated maximal probability of failure to
nearly one, without observing an actual breakdown of the algorithm.
\section{Implementation and Tests}
\label{sec-3}
\label{SEC:tests} To demonstrate the effectiveness of our algorithm,
we have written an implementation in
Haskell~\cite{peytonjones-haskell-98-language-2003}.  The close
resemblance of its terse syntax to mathematical notation, along with
its flexible type system makes it very convenient to express
algorithms in this language\footnote{Interesting applications of Haskell
to scientific computing can be found, for example,
in~\cite{Karczmarczuk:1997:GPL:270783.270796,karczmarczuk1999scientific,lobachev2008towards}.}.
We call our program ICE, the \emph{IBP Chooser of Equations}.  It is
available for downloading at \url{http://www.physik.hu-berlin.de/pep/tools}.

In order to test our algorithm, we generate equations for some
well-studied but non-trivial diagrams and filter them with our
program.  The generation of the equations has been performed with two
codes, CRUSHER by Peter Marquard, and a program under development by
Johann Usovitsch.  

\begin{figure}[htb]
\centering
\includegraphics[width=.5\linewidth]{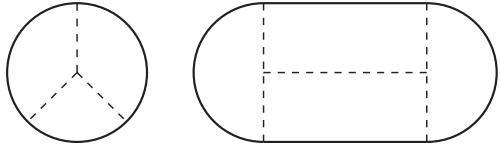}
\caption{\label{fig:dias}The three- and four-loop vacuum topologies BM (left) and H (right). The solid lines are massive, the dashed are massless.  The names follow the notations in~\cite{Broadhurst:1991fi,Avdeev:1995eu,Steinhauser:2000ry,Schroder:2005va}.}
\end{figure}

The diagrams used in our tests are depicted in
Figure~\ref{fig:dias}.  We generate equations to reduce integrals
with up to a certain number of \emph{dots} (additional powers of
propagators).  We note the number of generated equations and the
number of equations our program selects as linearly independent.  We
check our results by observing that the master integrals found by ICE
coincide with those in the
literature~\cite{Broadhurst:1991fi,Avdeev:1995eu,Steinhauser:2000ry,Schroder:2005va}. The
results are shown in the table below.  Depending on the individual
system, we see that about one half up to three quarters of the
equations are eliminated.

\begin{center}
\begin{tabular}{lrrrr}
Topology & Dots & Equations & Independent Equations & Ratio\\
\hline
H & 1 & 10464 & 5767 & 0.55\\
H & 2 & 39600 & 18626 & 0.47\\
BM & 3 & 3114 & 1148 & 0.37\\
BM & 10 & 113571 & 28851 & 0.25\\
\end{tabular}
\end{center}

\subsection{Selecting Specific Equations}
\label{sec-3-1}
While the number of linearly independent equations is fixed for a
given system, there is some arbitrariness in the choice of which
equations to keep.  Depending on the specifics of the implementation
of Laporta's algorithm, a clever selection can have a great impact on
the running time of the reduction.
In ICE, we try to minimise the number of entries below the diagonal
of the resulting system, bringing the system as close to an upper
triangular form as possible before solving any equations.
\subsection{Optional Backwards Elimination}
\label{sec-3-2}
We have implemented the possibility to perform not only a forward, but
also a backwards elimination.  This allows to determine which master
integrals appear in the reduction of each scalar integral.

\section{Conclusions}
\label{sec-4}
The computational cost of Laporta's algorithm is driven by two
inconveniences: intermediate expression swell and a large amount of
redundant information in the input.  We have described a simple
algorithm to deal with the latter problem by selecting a maximal
linearly independent set of equations from the input.  It is a Monte
Carlo algorithm, i.e., it has deterministic running time and gives the
correct answer with high probability.  In particular, the cost of
running the algorithm is virtually independent of the number of
kinematic invariants in the problem:  after evaluating the initial
IBP equations at a random point, all arithmetic is performed in a
finite field.

Our algorithm determines, en passant, a set of master integrals needed
for the reduction of a specific class of diagrams.  With some
additional effort, it is also possible to determine which master
integrals will appear in the result for specific integrals.

Recently, there has been some work on orthogonal methods to determine
the set of master integrals without performing a full
reduction~\cite{Lee:2013hzt,Robloopfesttalk}.  With such
algorithms, it should be possible to guarantee success of our
algorithm (at the possible expense of additional computing time due to
repeated runs) by comparing the identified master integrals with their
predicted number.  In other words, the methods
of~\cite{Lee:2013hzt,Robloopfesttalk} deliver a criterion for the
correctness of the result of our algorithm, so that it can be turned
from a Monte Carlo to a Las Vegas algorithm.

\section{Acknowledgements}
\label{sec-5}
We thank Peter Marquard and Johann Usovich for help in testing the
algorithm, and Elisabeth Kant, Robert Schabinger, Matthias
Steinhauser, A.V. and V.A. Smirnov, J. Bas Tausk, and Peter Uwer for
valuable comments on the draft of this paper.  This work was supported
by the DFG through SFB/TR9.

\bibliographystyle{elsarticle-num}
\bibliography{ibp-lindep}
\end{document}